\documentclass[manuscript]{acmart}

\AtBeginDocument{%
  \providecommand\BibTeX{{%
    \normalfont B\kern-0.5em{\scshape i\kern-0.25em b}\kern-0.8em\TeX}}}

\setcopyright{acmcopyright}
\copyrightyear{2023}
\acmYear{2023}
\acmDOI{XXXXXXX.XXXXXXX}

\acmConference[CHI 2023]{Make sure to enter the correct
  conference title from your rights confirmation emai}{April 23--28,
  2023}{Hamburg, Germany}
\acmBooktitle{CHi 2023:  ACM CHI Conference on Human Factors in Computing Systems ,
 April 23--28, 2023, Hamburg, Germany} 
\acmPrice{15.00}
\acmISBN{978-1-4503-XXXX-X/18/06}

\begin{document}

\title[Explainable Human-Robot Training and Cooperation through AR]{Explainable Human-Robot Training and Cooperation with Augmented Reality}

\author{Chao Wang}
\authornote{All authors contributed equally to this research.}
\email{chao.wang@honda-ri.de}

\author{Anna Belardinelli}
\authornotemark[1]

\author{Stephan Hasler}
\authornotemark[1]

\author{Theodoros Stouraitis}
\authornotemark[1]

\author{Daniel Tanneberg}
\authornotemark[1]

\author{Michael Gienger}
\authornotemark[1]

\affiliation{%
  \institution{Honda Research Institute EU}
  \city{Offenbach am Main}
  \state{Hessen}
  \country{Germany}
}



\renewcommand{\shortauthors}{Wang, et al.}

\begin{abstract}
The current spread of social and assistive robotics applications is increasingly highlighting the need for robots that can be easily taught and interacted with, even by users with no technical background. 
Still, it is often difficult to grasp what such robots know or to assess if a correct representation of the task is being formed. Augmented Reality (AR) has the potential to bridge this gap. We demonstrate three use cases where AR design elements enhance the explainability and efficiency of human-robot interaction: 1) a human teaching a robot some simple kitchen tasks by demonstration, 2) the robot showing its plan for solving novel tasks in AR to a human for validation, and 3) a robot communicating its intentions via AR while assisting people with limited mobility during daily activities.
\end{abstract}




\begin{teaserfigure}
  \includegraphics[width=\textwidth]{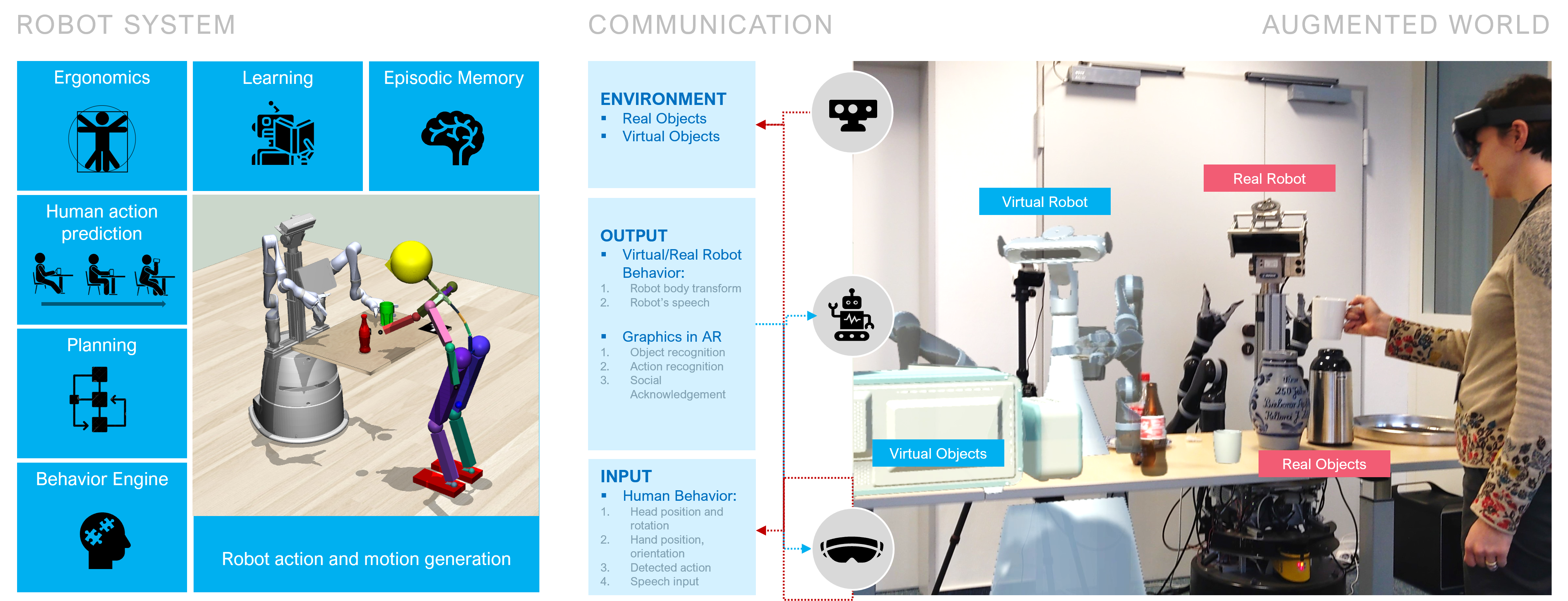}
  \vspace{-4mm}
  \caption{System architecture for AR-mediated training and cooperation with robots.}
  \label{fig:teaser}
\end{teaserfigure}

\received{19 January 2023}

\maketitle

\section{Introduction}
Among the foremost conditions to have socially assistive robots enter our homes and act around us, is the possibility even for non-expert users to be able to intuitively interact with the robot and to teach it new tasks.
Such systems are meant for long-term interaction, hence they will need to be flexible enough to handle a large number of tasks and to learn new ones. This means that their functioning will be rather complex, relying on multiple AI and machine learning algorithms which cannot be fathomed by the end user. At the same time, robots will need to inform the user about their workings in an effective way, so that the user can adjust its mental model and understand what to expect from the robot without the need for specific training, lengthy instructions, or repeated observations of the robot. Such a feature would go a long way in improving acceptance and trust in robotic assistants by the public. This need, already emerged in recent years in disembodied AI systems and tackled in the eXplainable AI (XAI) field \cite{gunning2019}, has been previously denoted as interpretability or legibility in Mobile Robotics \cite{Dragan_2013}, and has been recently characterized by specific robotic dimensions \cite{Hellstroem_2018}. Indeed, while pure AI systems usually provide an a-posteriori explanation on decisions/predictions made by some machine learning algorithm, robots are embodied agents, autonomously acting in the real world and possibly interacting and cooperating with humans. This situatedness and agency calls for different kinds of explanations, not just regarding past decisions, but more critically about current assessments, future plans, and intentions. Such information needs to be timely communicated to the human user and interactively negotiated with them in a quick and unambiguous fashion. In Human-Robot Interaction (HRI) similar interactions have often been managed by imitating interactions and social cues exchanged between humans, e.g., gaze contact and shared attention, speech, gestures \cite{Wallkoetter2021}, with the idea that behavioral human-likeness can improve intuitiveness. Still, due to their familiarity, these cues might falsely suggest that the robot has similar perceptual and understanding capabilities as the human counterpart. In this sense, other interaction modalities might integrate these cues and offer an insight into the robot’s mind that helps bridging the user’s mental model gaps. 

We propose a concept utilizing AR to improve the user experience of interacting with robot assistants by displaying intuitive explanatory hints about the robot perception, learning, generalization, and planning at different levels. While Augmented/Mixed Reality solutions have been recently increasingly spreading in robotics \cite{Bassyouni_2021} 
, these have typically targeted debugging by trained experts \cite{Rotsidis_2019}, visualizing internal states and movement intentions \cite{Walker2018},
teleoperation \cite{yew2017}, and simplifying learning by demonstration of movement trajectories \cite{luebbers2019}. 
Here, we introduce our robotic system which leverages AR capabilities across different typical use cases in robotics: learning new tasks by demonstration, devising a plan for a new problem, and assisting a human during a physical task in an ergonomic way. By exploiting real and virtual objects and by integrating human-like social cues with explanatory virtual design elements \cite{Walker2022}, this framework showcases how a hybrid workspace can be shared between humans and robots. Such interfaces enhance the interaction by making the robot's state of mind visually apparent and transparent to the user, grounding it on the current human perception of the scene.



\section{System Architecture}
Our system is composed of modules realizing the back-end functionalities (learning, planning, prediction and motion generation), and the front-end interface capabilities (visualization and interaction) of the AR glasses (see Fig. \ref{fig:teaser}). 

\textbf{Back-end robotic system}: The physical robot is a torso with two Kinova arms\footnote{\url{https://www.kinovarobotics.com/product/gen2-robots}} with 7 DOF each and a  pan-tilt unit, part of the mobile platform "Johnny" \cite{Wang2021}. 
The robot's cognitive skills are realized in multiple ROS\footnote{\url{https://www.ros.org}} nodes, communicating with each other and with the front-end interface. The \textit{behavior engine} orchestrates the social behavior of the robot. It controls the robot gaze, while concurrently issuing XAI cues to be shown in the AR environment (cfr. \cite{wang2022}). It also regulates the speech interaction, e.g. acknowledging the user commands or asking curiosity-driven questions (see Sec. \ref{sec:usecase1}). The \textit{episodic memory} collects world state observations during demonstrations by the human teachers. 
The \textit{learning} module realizes symbolic skill learning which integrates demonstrations from the episodic memory into a knowledge graph, and generates new hypotheses to be queried to the user (Sec. \ref{sec:usecase1}). The \textit{planning} module uses the learned semantic skills to generate high-level plans to solve new tasks. Such plans are yet to be checked with the human tutor (see Sec. \ref{sec:usecase2}). The \textit{human action prediction} module operates when assisting a user. The robot can predict what the user is intending to achieve, and plan a supportive action, e.g., moving the next required object closer to the user. The \textit{ergonomics} module generates interventions in a ergonomically optimal way for the user (see Sec. \ref{sec:usecase3}). Finally, the \textit{action generation} module (Fig. \ref{fig:teaser}) receives action commands from other modules (e.g., where to look, what to grasp, etc) and executes the corresponding motor behavior.

\textbf{Front-end interface}: Virtual objects, a virtual robot, and XAI cues (AR graphical elements) are displayed in the mixed-reality environment via the HoloLens\footnote{\url{https://www.microsoft.com/en-us/hololens}}. The HoloLens can scan the surroundings, build up 3D meshes of the environment objects and locate itself in the room, which enables it to stably overlay graphics in the environment considering occlusions with real objects. \par
The shared environment, accessible to the user through the HoloLens (see Fig.  \ref{fig:teaser}, right), includes multiple real and virtual objects. The object poses are continuously sent to the back-end, by the HoloLens for virtual objects and by a static camera for real objects (endowed with fiducial markers). 
Users can interact with the virtual objects in a similar way as with real ones; they can, for instance, pick up the bread, put it into a slot of the toaster, and push the button. \par
The HoloLens is also detecting the user's behavior and communicates it to the back-end system. This includes the user's head position, orientation and speech input. More importantly, as the HoloLens can track the user's hand and fingers, then the manual actions (e.g., "pick" or "drop") are also detected. The manipulation of objects by a human hand is implemented via Microsoft MRTK SDK\footnote{\url{https://learn.microsoft.com/en-us/windows/mixed-reality/mrtk-unity/mrtk2/?view=mrtkunity-2021-05}}, which enables the corresponding virtual object to stick to the human's hand while the "picking/holding" gesture is applied, and release from the hand after a "drop" gesture is detected. Moreover, colliders are attached to the user's fingers, enabling the teacher, for instance, to press the toaster lever, turn on the power button of the microwave and close the microwave door.
Finally, the user behavior and related manipulated object information are sent to the back-end system via ROS (see Fig. \ref{fig:teaser}, "INPUT").
The system can also display and animate a holographic virtual robot, which looks almost identical as the physical robot. 
The HoloLens receives the robot behavior data (see Fig. \ref{fig:teaser} "OUTPUT"), including the pose of the virtual robot and the speech commands, and visualizes / speaks them. Furthermore, the back-end  triggers the display of the XAI cues which are shown in the AR environment.
The next sections introduce three use cases for enhancing human-robot interaction via AR based on this system architecture. 

\section{Use case 1: Explainable human-robot interaction for imitation learning} \label{sec:usecase1}
\begin{figure}
    \centering
    \includegraphics[width=1\textwidth]{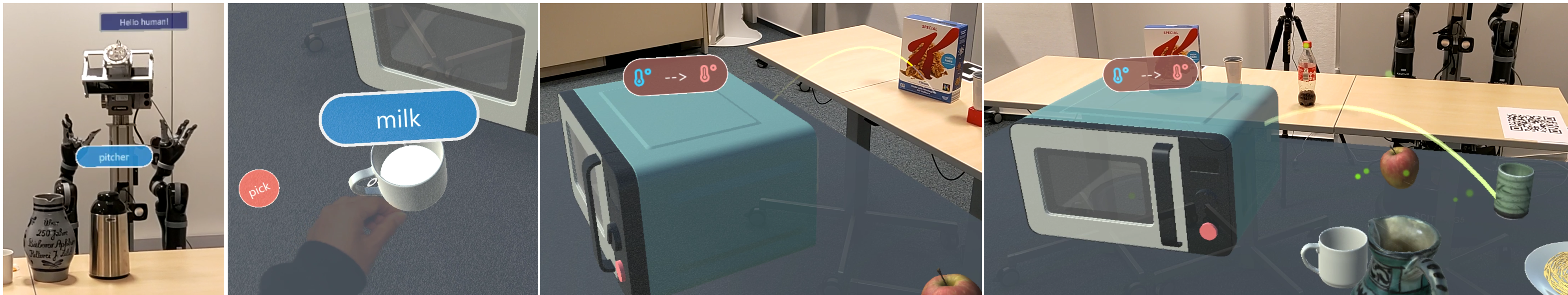}
    \vspace{-7mm}
    \caption{Learning from demonstration with XAI: while the user demonstrates a new skill the robot signals in AR which actions and objects it recognizes. Afterwards, it asks a related question while highlighting involved objects.}
    \label{fig:demo-1}
    \vspace{-4mm}
\end{figure}
Learning by demonstration or imitation learning in robotics aims at enabling humans to teach new skills by physically showing the task as they would do to another person. Major limitations in such an approach have been the capability of the robot to interpret and generalize a specific demonstration, and the difficulty that human users have to understand what constitutes a good demonstration for the robot. 
We therefore developed a two-stage skill learning concept. In the first stage, the user demonstrates a skill to the robot, which acquires it using semantic skill learning concepts. The learned representation of the skill is formed by symbols that encode preconditions, actions, and effects. In the second stage, the robot takes initiative and asks curious questions about the demonstrated task to the user. Both stages are designed to enhance the user's mental model of the system using AR and social cues.

Specifically, the user (wearing AR glasses) demonstrates a skill to the robot, e.g., by interacting with the (virtual) objects and giving language explanations (skill labels). The robot follows the teacher's gaze or looks back at him to show its attention. The teacher is further continuously informed about the robot's perception: Object and action labels (XAI cues) are popping up whenever the teacher gazes at some object or a manual action is recognized (cfr. Fig. \ref{fig:demo-1} and \cite{wang2022}).
The state of the environment, including the objects and agents, is recorded before and after the demonstrated skill by the episodic memory.
These observations are parsed into predefined symbolic representations of the environment, consisting of logical predicates.
Together with the skill label, these observations are used to learn the symbolic skill, capturing in which situations it can be applied and what is changed in the environment by executing it.

If skills with similar effect but different objects are learned, the robot can generalize these skills using a predefined object hierarchy. This type-generalization of skills enables to consider novel objects in a new task.
This symbolic nature of the skill representation enables the robot to 
explain the learned knowledge or a future plan in an intuitive way.

While such an imitation learning scenario is intuitive and engaging for the user, it still takes time and effort to demonstrate the skills. 
To speed up the interactive learning process, the robot has been equipped with the ability to ask curious questions about skills and objects, i.e., creating hypotheses that can be presented to the user for confirmation or rejection.
This allows teaching additional knowledge to the system in an interactive and, in comparison to full skill demonstrations, faster way. For example, after seeing a demonstration to learn to use the 
microwave, where the teacher heated milk, the robot can ask whether a similar object (e.g., water) can be heated in the microwave. The XAI virtual elements in this case would highlight the candidate object and the microwave and, after the user's answer, acknoweledge with red/green particles the negative/positive answer (see Fig. \ref{fig:demo-1}, right panels).


\section{Use Case 2: Visualization of the robot planning} \label{sec:usecase2}
Learned skills contain both high-level knowledge about the physical effects of different devices and low-level knowledge on how to operate these devices and manipulate objects.
We use a standard symbolic STRIPS planner to combine the skills to solve novel and more complex tasks considering present objects, both real and virtual ones. For example, when asked to 'Prepare an ice tea' the planner might consider the kettle or the microwave to heat some water before putting a tea bag inside and the fridge or some ice cubes to cool it later. 
Such a high-level solution is extended with the required low-level manipulations, like placing objects, opening doors and pressing buttons. 

\begin{figure}
    \centering
    \includegraphics[width=1\textwidth]{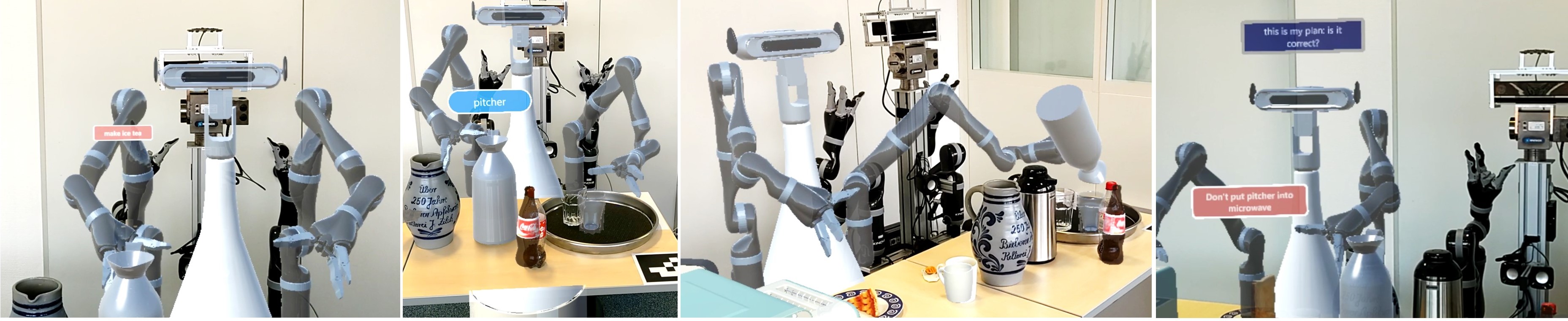}
    \vspace{-7mm}
    \caption{Visualization of planning phases, from left to right: the user instructs the robot by speech; the robot shows its avatar and digital twins of real objects; the avatar executes the plan; the human gives feedback on the plan. Red boxes show the user's commands as detected by speech recognition.}
    \label{fig:Planning-1}
    \vspace{-4mm}
\end{figure}

By learning the skills and applying them to new tasks and environments, the system generalizes from previously observed episodes. As a consequence, the generated plans may not be feasible or desired and need validation by a human. Here, we propose an interactive system which can show the plan of the robot to the human via AR glasses before it is executed (see Fig. \ref{fig:Planning-1}). 
 The user can give a command to the robot via speech
 . Then the robot will generate a plan to solve the query according to its knowledge at different levels. Before execution, the robot will ask the user to validate the plan in AR. The virtual "avatar" of the robot appears overlaid on the physical body of the robot and real object "shadows" (holographic twins) are displayed in the AR glasses. Then the virtual robot will execute the plan with the virtual objects.
 In this way, the human can understand the robot reasoning and provide feedback to its plan, approving or correcting it.

\vspace{-2mm}
\section{Use case 3: Communicating robot's intentions while assisting users}\label{sec:usecase3}

\begin{figure}
    \centering
    \includegraphics[width=1\textwidth]{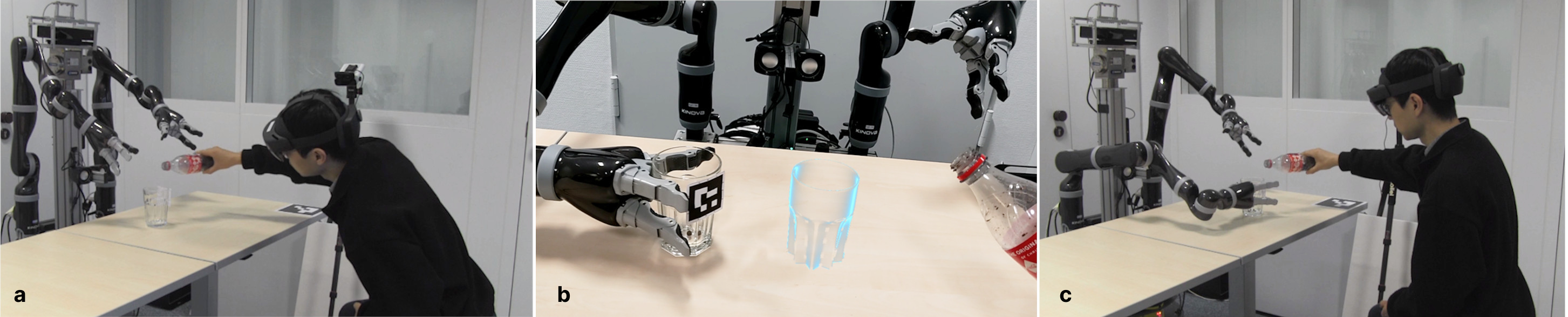}
    \vspace{-7mm}
    \caption{Instance (a) occurs in a scenario without assistance, where the human bends to pour into the glass, resulting in a inconvenient posture. In contrast, instances (b) and (c) are two consecutive frames from a scenario with assistance. In (b), the robots predicts that the human will pour into the glass and reveals the outcome of its intervention utilising XAI cues while relocating the glass. In (c), the human pours into the glass in a comfortable posture.}
    \label{fig:ergonomics-1}
    \vspace{-4mm}
\end{figure}

Efficient cooperation between a human and a robot includes a number of  challenges. These, among others, are: (i) human behaviour anticipation,
(ii) predicting the influence of the robot's actions onto the human's actions,
and (iii) generating legible robot actions~\cite{Dragan_2013} that can be easily understood by the human partner. Here, we study these challenges in the context of manipulation where a robot assists a human to perform a sequence of actions. As a typical example, consider a human (sitting on a wheelchair) serving a beverage, as shown in Figure~\ref{fig:ergonomics-1}. This scenario requires the human to bend/stretch to reach the glass, hence, a robot assistant could be of aid. 


To support the human, the robot ideally predicts the sequence of likely human actions and postures~\cite{StouraitisTRO2020}, assesses the future human postures~\cite{van2020predicting}, computes the desired intervention that improves the physical state of the human, and communicates the effects of the intended interventions to the human ahead of time. 
To achieve these aims, we propose a model-based optimisation approach that performs a physics-based prediction of the human's actions and enables the robot to decide which intervention will improve the human posture, hence their ergonomic state, and illustrate them to the human via the XAI cues. 


In this use case we consider a table-top scenario, where the location of several objects and the intended high-level task is provided to the system. Such a task can be serving a beverage or a bowl of cereal. 
Our system is able to: (i) 
predict the sequence of humans actions, e.g. picking up a bottle and then pouring into a glass. (ii) Evaluating the human posture, e.g. human upper-body configuration while pouring, and decide how to adjust continuous quantities, e.g. change the pose of the glass to improve the human's upper-body configuration while pouring.  
(iii) Inform the human using AR (e.g. using holograms to show where objects will be relocated), while performing the assistive action. Utilizing AR to reveal the outcome of a future invention allows the user to understand the robot actions, enabling him to comfortably and fluently perform the task. 

\section{Conclusions}
Our system aims at providing critical information about the robot, alleviating the cognitive and physical load of the user who can act and interact in a natural way, without constantly monitoring and scrutinizing the robot but correcting it only if necessary. While we target here the specific case of robots assisting in a domestic environment, we implement AR-based design concepts that could cater to understanding and justification needs increasingly present in the interaction with physical intelligent systems and hopefully can inspire the broader CHI community with related solutions for smart homes or other human-centered AI systems. Still, the focus of our work is on facilitating robot interaction for non-experts and on increasing the independence of elderly and people with limited mobility, thus creating an impact on accessibility and inclusiveness in human-machine interaction.




\bibliographystyle{ACM-Reference-Format}
\bibliography{xhri_bib}






\end{document}